# A novel dielectric elastomer actuator based on polyvinyl alcohol hydrogel compliant electrodes

Chengyi Xu, Chunye Xu, Jianming Zheng*

Department of Polymer Science and Engineering, Hefei National Laboratory for Physical Sciences at the Microscale, University of Science and Technology of China, Hefei 230026, P.R. China

## ABSTRACT

We firstly demonstrate physically-prepared compliant PVA hydrogel electrodes as a promising supplement in dielectric elastomer actuators. They are capable of conducting high voltages over 5 kV without electrolysis by an electrical double layer effect. The hydrogel electrodes adhere tightly to the dielectric layer, which ensures its structural stability during actuation. All-polymeric actuators fabricated by these electrodes exhibited excellent consecutive working performance over 2960 cycles. The actuation was influenced by applied voltage, ramp rate, electrode elastic modulus and moisture content. A maximum areal strain over 78% was achieved. Tunable transparency, good biocompatibility, long lifetime, low cost and facile fabrication make PVA hydrogel electrode another promising candidate in the fields of sensors, artificial muscles and optical applications.

**Keywords:** dielectric elastomer, hydrogel electrode, maximum strain*

## 1. INTRODUCTION

Recently, electroactive polymers (EAPs) have emerged as a new class of intelligent materials. They can reversibly convert electrical energy into mechanical work by changing shape or size in response to external stimuli[1]. EAPs share numerous merits such as low density, low cost and easy operation[1-2]. These unique characteristics make EAPs of great value in the fields of robotics[3], sensors[4], optical electronics[5] and microelectromechanical systems[6]. Among these, dielectric elastomer (DE) has been regarded as a promising candidate in the realization of artificial muscles[3, 7], owing to its flexibility, instant response and huge deformation.

Dielectric elastomer actuator (DEA) consists of a soft elastomer (usually electrical insulator) sandwiched by a pair of compliant electrodes. When a high voltage is applied, opposite charges aggregates at different electrodes, leading to the formation of an electrostatic force. The electrostatic force compresses the elastomer longitudinally and thus causes substantial planar deformation. For an ideal monolayer DEA, the deformation ($D$) across the thickness is expressed as Equation (1), where $\varepsilon_0$ and $\varepsilon_{DE}$ are respectively the dielectric constant of vacuum and dielectric elastomer, $Y$ is the Young's modulus, and $V/z$ is the field strength.

$$D = \frac{1}{Y}\varepsilon_0\varepsilon_{DE}\left(\frac{V}{z}\right) \qquad (1)$$

To enhance the actuation performance, increasing the field strength is the most direct and convenient method. Higher electric strength can be achieved by increasing the applied voltage across the DEA, or prestraining the dielectric on the order from 200% to 500%. So far, a variety of materials such as acrylates[8], silicones[9] and polyurethanes[10], have been reported to achieve actuation as dielectric layers. Much work has been done to improve the dielectric constant of DE by means of functionalizing polar groups[11] (e.g. trifluoromethyl, acetylamino) onto polymer skeleton, forming co-polymers with materials of strong polarity[12] (e.g. polyacrylamide) or blending fillers[13] (e.g. copper(Ⅱ) phthalocyanine, $\varepsilon \approx 50,000$) as multi-phase composites. Meanwhile, the realization of giant deformation relies on conductive, flexible compliant electrodes with Young's modulus lower than the DE[7]. Carbon grease[14], graphite spray[15], metal nanowires[16] and conductive oxides[17] are common candidates as electrode materials. However, these electrodes suffers from poor structural integrity during the cyclic process.

* E-mail: jmz@ustc.edu.cn; Tel &Fax: +86-551-360-3470; http://www.hfnl.ustc.edu.cn/2009/0605/955.html
  Address: Hefei National Laboratory for Physical Sciences at the Microscale, University of Science and Technology of China, Hefei, 230026, P. R. China

To overcome those limitations, polyacrylamide (PAM) electrodes was reported by C. Keplinger in 2013, which were capable of conducting kilovolt-level voltage without electrochemical degradation[18]. However, the preparation of PAM electrodes required to undergo irradiation polymerization with peroxides and crosslinking agents. The residual monomer and degradation products have known toxic effects on human nervous system. For bio-inspired applications, here we firstly introduce an all-polymeric dielectric elastomer actuator based on complaint polyvinyl alcohol (PVA) hydrogel electrodes with excellent performance. Traditional electrodes made of metallic or carbon films were replaced by physically-treated transparent ionic gels. In addition to good biocompatibility for medical use, PVA electrodes can be easily prepared with flexible networks through inter and intramolecular hydrogen bonds. This interaction makes the hydrogel electrodes stretchable and electroactive even after 2,500 cycles. Moreover, the soft PVA hydrogel electrodes adheres well to the dielectric layer to prevent the slippage of interfaces, and can achieve large deformation over 78%.

## 2. EXPERIMENTAL SECTION

Lithium chloride (Sinopharm Reagents, China) was dissolved into deionized water. Then commercial PVA (Kuraray 117, Japan) powder was added into the solution. The mixture was vigorously mechanical stirred for one hour at 90℃. After PVA was completely dispersed, the obtained transparent and homogeneous liquid of high viscosity was immediately transferred into a PTFE Petri Dish mold. In order to facilitate the formation of physical crosslink, repetitive cycles of freezing (-20℃, 23 hours) and thawing (room temperature, 1 hour) were carried out for five times. The finally obtained hydrogel film was peeled off from substrate and cut into different sizes and shapes for further study.

A prototype of all-polymeric actuator was fabricated by using 3M VHB 4910 acrylate tape as dielectric elastomer. The Young's modulus was measured about 220 kPa. The DE layer was prestretched to 300% and fixed with a circle of plastic stiff frame. Then circular PVA hydrogel (1 mm thick) were laminated at both sides as a sandwiched structure. The hydrogels were directly connected to a power supply via conductive wires.

## 3. RESULTS AND DISCUSSION

Here we demonstrate PVA hydrogel electrodes as a promising supplement in DEA for its high stretchability, transparency and biocompatibility. The compliant hydrogel electrodes can deliver high voltages over 5 kV, which can be interpreted by an electrical double layer theory (Fig. 1(a)). In particular, ionic components, as $Li^+$ and $Cl^-$, transfer and rearrange directionally at the hydrogel-dielectric interface in response to electric field (Fig. 1(b)). Different signs of charges aggregate rapidly at the two interfaces and thus the Coulombic attraction causes a Maxwell stress to constrain the dielectric. The soft electrode hydrogel expands outward simultaneously owing to its good adhesion and low Young's modulus. The heart-beating pump is presented in the supplemental video. As the heart expands and contracts, a transparency shift was observed from 40 % to 72 %, clearly presented in Fig. 2(a) and (b).

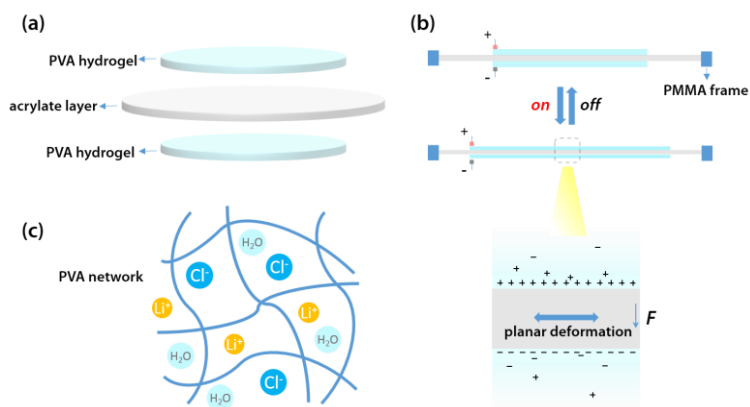

**Fig. 1. Schematic of dielectric elastomer actuator with compliant PVA hydrogel electrodes.** (a) A sandwiched structure. (b) Principle of reversible DEA actuation, and electrical double layer formed at the hydrogel-dielectric interfaces. (c) Illustration of ionic PVA network inside hydrogel electrodes.

We observed the electro-induced actuation of VHB film. To illustrate the response, deformations of two separate actuators were recorded by a high-speed camera (250 frames per second), as plotted in Fig. 2(e). One was fabricated with compliant PVA hydrogel electrodes, and the other was based on silver nanowire (AgNW) electrodes. During the first cycle, the hydrogel-based actuator achieved 95% of maximum strain in 0.72 s at 3.5 kV and recovered in 0.93 s after withdrawing the voltage. Meanwhile, AgNW-based actuator inherits excellent electron transportation ability of metallic materials. Reversible sliding nanowire networks enabled AgNW electrodes to show a faster response in 0.432 s at 3.5 kV, reaching 95% of maximum areal strain with its optical transmittance varying from 28% to 41% (Fig. 2(c) and (d)). The lags of hydrogel electrode in the response test are due to its intrinsic viscoelastic property, as well as the ion-driven mechanism in aqueous solvent. Both the two electrodes largely enhanced their performance after 100 consecutive cycles of actuation.

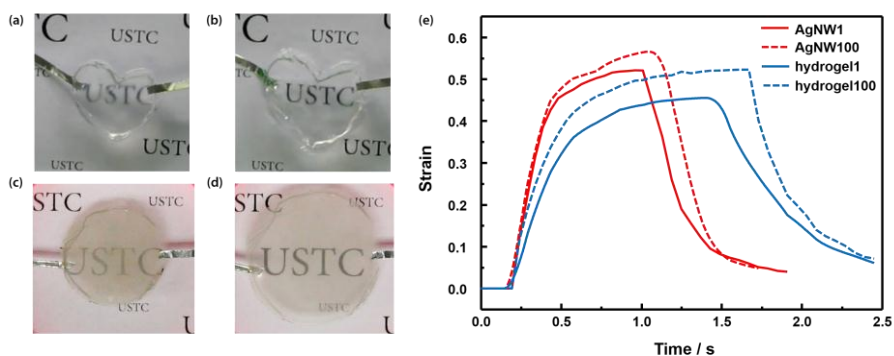

**Fig. 2. DEAs fabricated from hydrogel and AgNW compliant electrodes.** The initial (a) and activated (b) state of actuators based on heart-like PVA hydrogel. The initial (c) and activated (d) state of actuators based on circular

AgNW film electrodes. (e) Response time of PVA hydrogel and AgNW electrodes after 1 and 100 successive cycles respectively.

A strong-bonded polymeric network is formed between water and PVA chains by inter and intramolecular hydrogen bonds in the freezing-thawing treatment. The effect of hydrogel strength was investigated. Raw solution of 5% was too thin to form qualified sample. Hydrogels made from 8%, 10% and 15% PVA were successfully prepared, of which the Young's modulus were measured as 4.98 kPa, 9.59 kPa, 20.76 kPa respectively by a Dynamic Mechanical Analyzer (DMA Q800, TA Instruments, USA). The additional PVA acts as reinforcement to polymer backbone. Fig. 3(a) demonstrates that with the increase of PVA content, the maximum actuation exhibited a distinct decline. This decline is mainly attributed to the increasing elastic modulus of the electrodes, which indicates that softer electrodes are desired in large actuation.

In order to examine the actuating properties, we varied the input voltage from 1.5 kV to 5 kV at the ramp rate of 0.5 kV/s. The maximum area strain considerably raises with the increase of working voltage, identical to Equation 1. At a given voltage, the maximum area strain appears to slightly increase with cycle numbers. This phenomenon is even more obvious at higher voltages. Low frequency relaxation occurs during dynamic cycles, which is associated with bond and chain segments reorientation in highly entangled polymer networks. After the dielectric elastomer relaxes to equilibrium, the deformation tends to maintain at a stable rate. Fig. 3(b) presents that the hydrogel electrodes can achieve an areal strain greater than 78% at 5 kV, and worked with negligible hysteresis over successive cycles.

The influence of the rate of increasing voltage on actuation performance is significant as presented in Fig. 3(c). The areal strain achieved the maximum at different ramp speeds (from 0.5 kV/s to 2 V/s) to 5 kV. In fact, slow rate gives sufficient time for PVA hydrogel electrodes to undergo full stretch and recovery in every single cycle. This relaxation accumulates with time, rising from the viscoelastic property of our hydrogel electrodes with a low Young's modulus of about 4.98 kPa. Later in our research, the rate of 0.5 kV/s to 3.5 kV was set as default parameter for actuation.

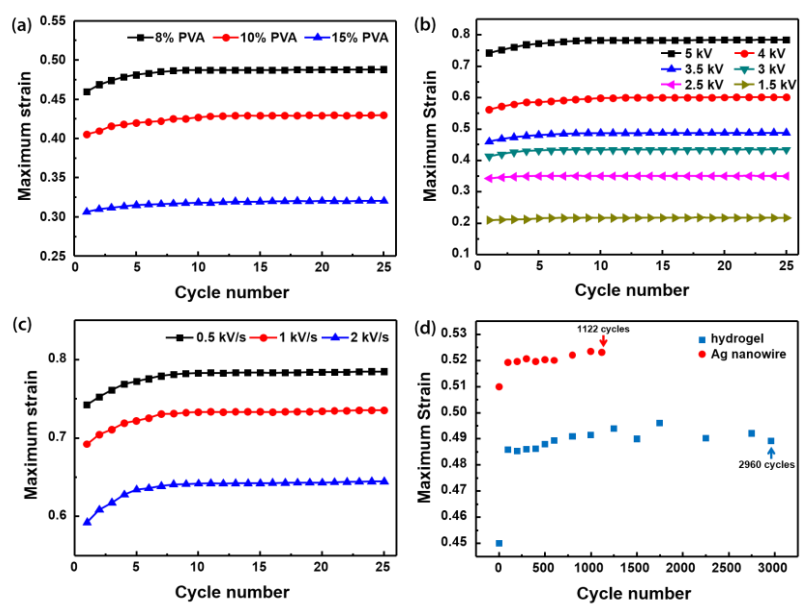

**Fig. 3. Actuation performance of PVA hydrogel electrodes.** (a) The influence of maximum strain under varied electric fields at a constant ramp speed of 0.5 kV/s over 25 successive cycles. (b) The influence of maximum strain on different ramp rates to 5 kV over 25 successive cycles. (c) The influence of PVA hydrogel strength on actuation performance. (d) Lifetime cyclic tests by PVA hydrogel and AgNW electrodes.

Lifetime is another key factor to measure continuous working performance. We took AgNW-based and gold-foil-based actuators as references, and recorded every 100 cycles. The gold actuator was constrained by poor flexibility and generated inconspicuous deformation less than 1%. Surface cracks were observed after 10 cycles for its brittleness. In comparison, both the AgNW and hydrogel electrodes approached stable maximum strains after first several hundred cycles due to the accumulation of chain relaxation capability (Fig. 3(d)). Particularly, the hydrogel-based actuator achieved 2960 consecutive cycles, almost three times as AgNW-based actuator. The intermediate dielectric layer finally broke down due to electric short. Failure or electrolysis of PVA hydrogels did not occur in our study even after exceeding the dielectric breakdown strength of VHB film, which ensures the reusage of PVA hydrogel electrodes by replacing the defective dielectric film.

The hydrogel electrodes remain electroactivity when stored at -25℃, however, they cannot be stored in air for long. Free water inside the polymer network suffers from a slow evaporation. We kept the hydrogel electrodes exposed at room temperature and 25% RH. Samples were exploited to perform actuation after 3, 7, 9, 15 and 24 hours respectively. The initial wet electrode was weighed as $M_0$. Then the tested sample was immediately dried by heating to measure the bulk weight of PVA as $M_1$. The moisture content was calculated to be 89.2%, 85.4%, 78.6%, 70%, 51.3% as follows:

$$\text{Moisture content} = (M_0-M_1)/M_0 \times 100\%$$

The actuator continued to generate large areal deformation for the first few hours. The margin area of hydrogel electrodes gradually turned curling and much stiffer with longer exposure time. Fig. 4 demonstrates that the maximum strain drastically decreased with the loss of water content. The electrodes completely ceased to work after 15 hours. For long-time storage in air, it is urgent to find efficient non-volatile electrolytes and develop encapsulation methods to prohibit dehydration.

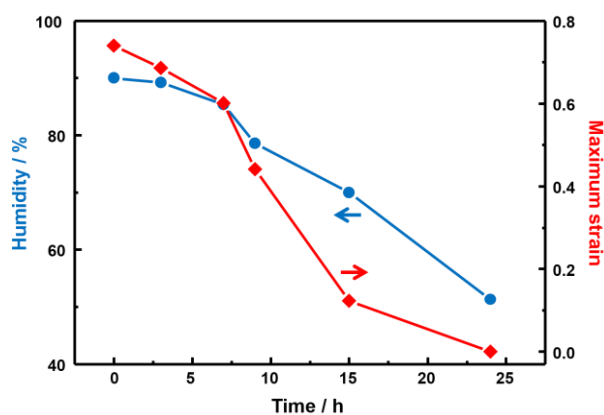

**Fig. 4.** The influence of moisture content on the electroactivity of hydrogel electrodes.

## 4. CONCLUSION

We firstly demonstrate physically-prepared compliant PVA hydrogel electrodes as a promising supplement in dielectric elastomer actuators. They are capable of conducting high voltages over 5 kV by an electrical double layer effect. The hydrogel electrodes adhere tightly to the dielectric layer, which ensures its structural stability during actuation. Actuators fabricated by these electrodes exhibited excellent consecutive working performance over 2960 cycles. The actuation was influenced by applied voltage, ramp rate and electrode elastic modulus. A maximum areal strain over 78% was achieved. Tunable transparency, good biocompatibility, long lifetime, low cost and facile fabrication make them another promising candidate in the fields of sensors, artificial muscles and optical applications.

## ACKNOWLEDGMENTS

This work received financial support from the National Natural Science Foundation of China (21273207, 21074125), Anhui Province Natural Science Foundation, China (11040606M57), the National Basic Research Program of China (2010CB934700), the "Hundred Talents Program" of CAS, and the "National Thousand Talents Program".